# Philipp Reis's telephone in Aberdeen

*John S. Reid & Rebecca Ronke*
*Natural Philosophy Collection, University of Aberdeen, Scotland*

**Abstract**

A telephone sender following the design of Johann Philipp Reis made by W. Ladd, London 1863, in the Natural Philosophy Collection of the University of Aberdeen is illustrated and described. How Reis came to invent a telephone some 15 years before Graham Bell's patent is discussed, with 19[th] century illustrations. The method employed by Reis is outlined and contemporary letters by Reis are reproduced, including an extensive letter to the instrument maker William Ladd. The origin of the Aberdeen instrument is given and some of the reasons why Bell was granted his patent in spite of the existence of Reis's instrument are suggested. The monument to Reis erected after his early death is shown.

**Introduction**

Alexander Graham Bell is so frequently quoted as the inventor of the telephone that it is interesting to find in our instrument collection a rare telephone sender that pre-dates Bell's early efforts by some 13 years. Undoubtedly it was paired with its receiver at one time, which may still come to light but it is more likely that being such a simple device the receiver has been disposed of without being recognised.

*Reis telephone sender by W. Ladd of London, 1863. The contacts on top of the membrane have been dislocated. They are not easily put back since the glass panel covering the membrane is held in place by a spring and is not readily removed. Inv. no. ABDNP203000*

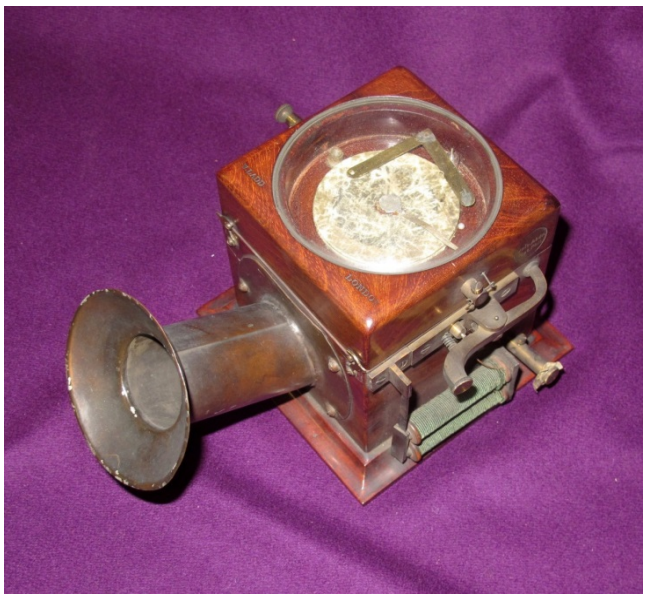
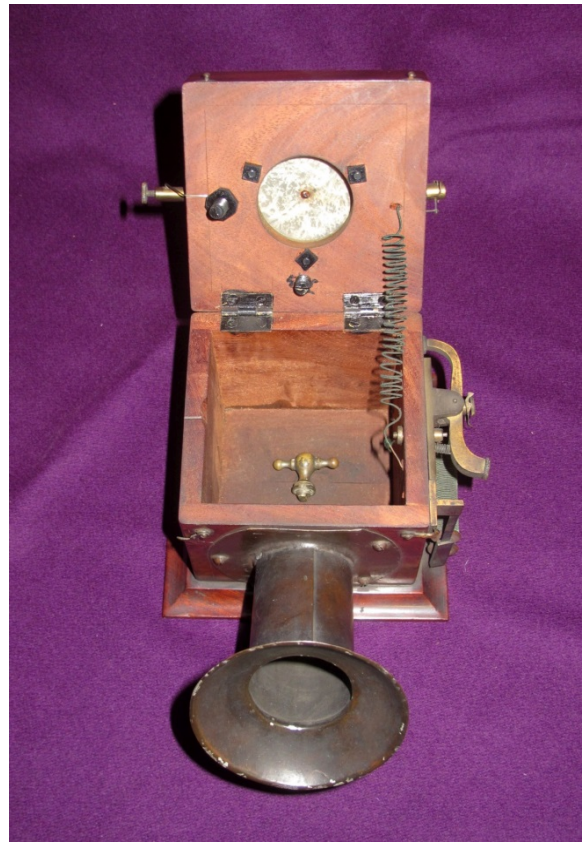

Whatever the American courts may have said at the time about Bell's 1876 and 1877 patents, Bell didn't invent the concept of the telephone, nor most of the technology that made it possible. The products of all the early inventors were seen as toys, including that of Bell. Bell, though, had the foresight to see beyond this label and with the help of backers he developed rapidly the commercial application of his version. Without Bell's drive, and perhaps the competition between Bell and Edison, the telephone may have taken much longer to become an 'everyday item' serving business and individuals around the world.

**Reis**

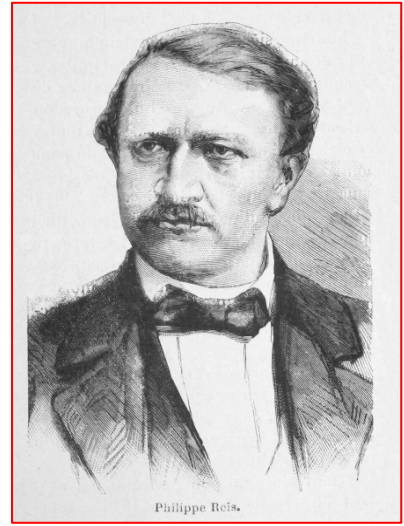

Perhaps the most unlucky of Bell's predecessors not to have his invention called the 'father of the telephone' was Johann Philipp Reis, 1834-1874. Many significant 19th century inventors were not professional scientists. Think of William Cooke, pioneer of the Cooke and Wheatstone telegraph, a young retired Indian army officer; Samuel Morse, an artist; or Bell himself, a professional teacher of deaf children.

Who was Philipp Reis? He was a well-educated German (adjacent figure[1]) who, on the instigation of his guardians, had learnt the trade of a paint dealer upon leaving school. As a hobby he attended classes and taught himself maths and physics. On the termination of his apprenticeship and his year of military service he left the paints business and took a teacher training course in Frankfurt. By 1859 he was a schoolmaster, teaching physical science at Garnier's Institute, a school he had attended in the 1840s in Friedrichsdorf, some 20 km from Frankfurt. It was here that he began experimenting with the electrical transmission of sound.

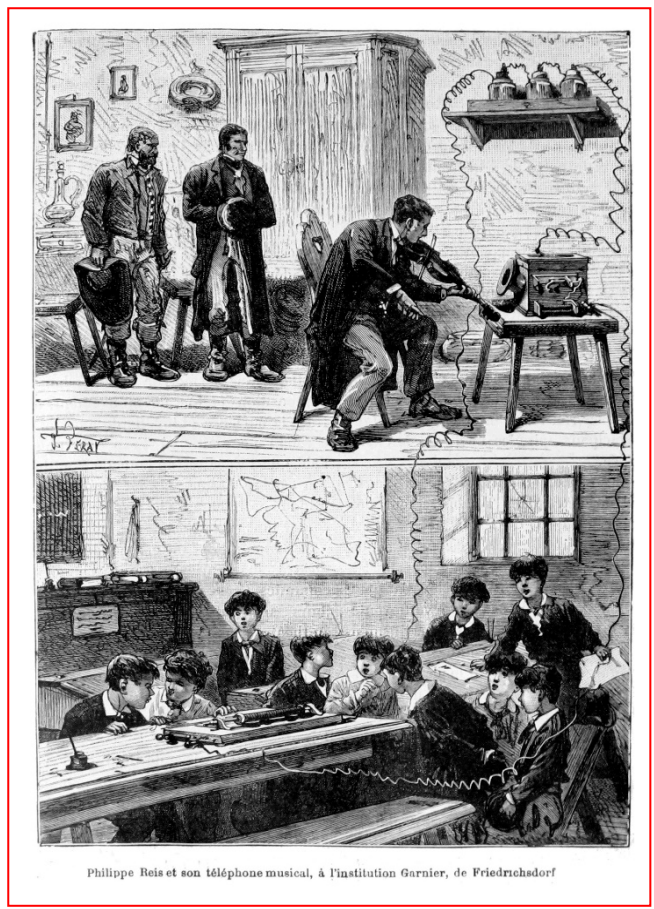

The adjacent print[2] is an artist's imagined view of Reis's telephone at his school. It is not wholly fantasy, for there is mention of some such arrangement by Sylvanus P. Thompson in his now rare but lengthy volume of 1883 on *Philipp Reis, Inventor of the Telephone: A Biographical Sketch*[3]. Thompson wrote: *He also carried a wire from the physical cabinet of Garnier's Institute across the playground into one of the classrooms for experimental telephonic communications; and a firmly established tradition of the school is still*

2/10

*preserved that the boys were afraid of making a noise in the class-room for fear that Herr Reis should hear them in his place among his favourite instruments.*

In one of the many American court cases of the 1880s it was said that two former pupils at Garnier's Institute testified *that this Reis thing did talk there; that Mr. Reis would stand in a building a hundred feet away, and talk to the transmitter, and they would all stand around the knitting needle receiver placed on the table; that Reis would read a book into the transmitter "and it was a kind of every-day exercise" he would read a book into it; and they could all hear the reading, standing, as they did, around the table.* The illustration shows Reis playing, not speaking.

**The Reis telephone**

Reis developed various prototypes, one of which was a transmitter in the shape of a human ear with embedded coil. By 1861 he had developed what he called a *telephon* and demonstrated it to the Frankfurt Physical Society in a lecture entitled "*Telephony by Means of the Galvanic Current*" and an account was published in the Society's yearbook[4]. Reis didn't patent his invention. His motivation is said by many to have been the transmission of music and song. It's true that in public demonstrations he showed his telephone transmitting songs and music, but so did Bell fifteen years later. Indeed, Bell has described how he was led to the transmission of sound by electricity through the problem of transmitting tones[5]. The illustration below shows the Reis telephone's components. They work like this.

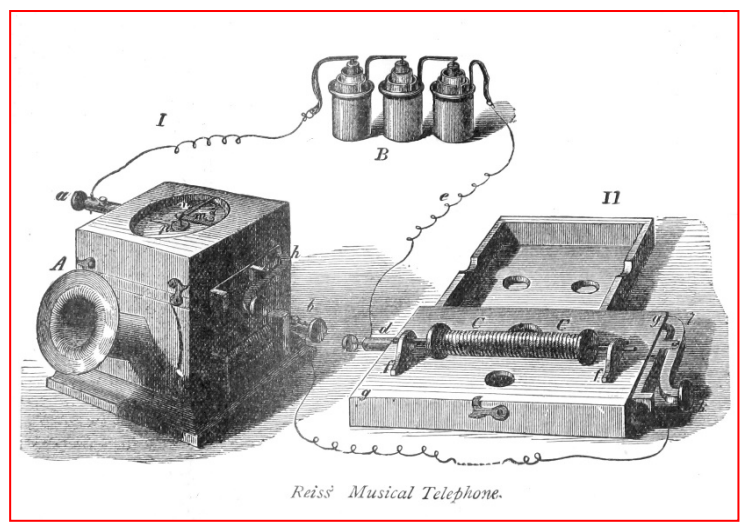

The transmitter design (on the left in the picture) was inspired by the construction of the human ear. Reis called it the 'singing station'. It uses the sound transmitted through the mouthpiece to vibrate a membrane on the top. On this membrane sits a delicate device to vary the contact of an electrical circuit with a battery, reminiscent of the how the cochlea is connected to the eardrum via three small bones. The receiver is a sounding box within which is a suitable magnetic rod (or needle) around which is wrapped a coil. The rod is secured to the wooden box. The varying current passing through the coil causes a varying magnetic field in the rod which, through the property of magnetostriction, causes small changes in its length in sympathy with the initial sound received by the singing station. These length changes result in impulses being given to the sounding box which thus repeats the sound received by the 'microphone'. If someone played a pure note A, for example, the membrane vibrated at 440 Hz and the battery circuit was interrupted at this frequency. Correspondingly the receiver sounded an A.

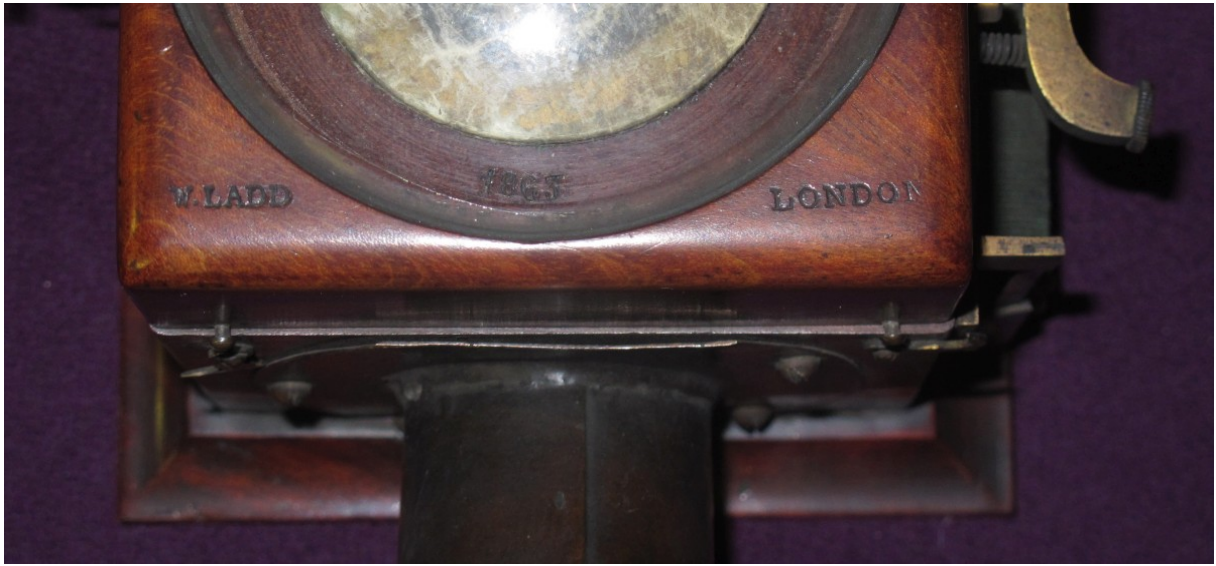

**Letters**

In the close-up of the top of our instrument '1863' is stamped under the glass. William Ladd, a London instrument maker who manufactured our specimen, purchased one of Reis's instruments from J. W. Albert & Son of Frankfurt in 1863. Reis subsequently wrote to Ladd in English. Since our example is made by Ladd, it is appropriate to quote Reis's letter in full with, I believe, original spelling from the copy included in Thompson's biography[6].

*Institut Garnier, Friedrichsdorf.*

*DEAR SIR!*

*I am very sorry not to have been in Frankfurt when you were there at Mr. Albert's, by whom I have been informed that you have purchased one of my newly-invented instruments (Telephons). Though I will do all in my power to give you the most ample explanations on the subject, I am sure that personal communication would have been preferable; specially as I was told, that you will show the apparatus at your next scientific meeting and thus introduce the apparatus in your country.*

*Tunes* [Reis means tones] *and sounds of any kind are only brought to our conception by the condensations and rarefactions of air or any other medium in which we may find ourselves. By every condensation the tympanum of our ear is pressed inwards, by every rarefaction it is pressed outward and thus the tympanum performs oscillations like a pendulum. The smaller or greater number of the oscillations made in a second gives us by help of the small bones in our ear and the auditory nerve the idea of a higher or lower tune.*

*It was no hard labour, either to imagine that any other membrane besides that of our ear, could be brought to make similar oscillations, if spanned in a proper manner and if taken in good proportions, or to make use of these oscillations for the interruption of a galvanic current.*

*However these were the principles wich (sic) guided me in my invention. They were sufficient to induce me to try the reproduction of tunes at any distance. It would be long to relate all the*

*fruitless attempts, I made, until I found out the proportions of the instrument and the necessary tension of the membrane. The apparatus you have bought, is now, what may be found most simple, and works without failling when arranged carefully in the following manner.*

*The apparatus consists of two separated parts; one for the singing station A, and the other for the hearing station B. [See the illustration at the end of the letter].*

*The apparatus A, a square box of wood, the cover of which shows the membrane (c) on the outside, under glass. In the middle of the latter is fixed a small platina plate to which a flattened copper wire is soldered on purpose to conduct the galvanic current. Within the circle you will further remark two screws. One of them is terminated by a little pit in which you put a little drop of quiksilver; the other is pointed. The angle, which you find lying on the membrane, is to be placed according to the letters, with the little whole* [hole] *(a) on the point (a) the little platina foot (b) into the quicksilver screw, the other platina foot will then come on the platina plate in the middle of the membrane.*

*The galvanic current coming from the battery (which I compose generally of three or four good elements) is introduced at the conducting screw near (b) wherefrom it proceeds to the quicksilver, the movable angle, the platina plate and the complementary telegraph to the conducting screw (s).* [This was the little auxiliary signalling apparatus at the side of the box, placed there for the same reasons as the auxiliary call-bell attached to modern telephones.]

*From here it goes through the conductor to the other station B and from there returns to the battery.*

*The apparatus B, a sonorous box on the cover of which is placed the wire-spiral with the steel axis, wich will be magnetic when the current goes through the spiral. A second little box is fixed on the first one, and laid down on the steel axis to increase the intensity of the reproduced sounds. On the small side of the lower box you will find the correspondent part of the complementary telegraph.*

*If a person sing at the station A, in the tube (x) the vibrations of air will pass into the box and move the membrane above; thereby the platina foot (c) of the movable angle will be lifted up and will thus open the stream at every condensation of air in the box. The stream will be re-established at every rarefaction. For this manner the steel axis at station B will be magnetic once for every full vibration; and as magnetism never enters nor leaves a metal without disturbing the equilibrium of the atoms, the steel-axis at station B must repeat the vibrations at station A and thus reproduce the sounds which caused them.*

<u>*Any*</u> *sound will be reproduced, if strong enough to set the membrane in motion.*

*The little telegraph, which you will find on the side of the apparatus is very usefull and agreable for to give signals between both of the correspondents. At every opening of the stream and next following shutting the station A will hear a little clap produced by the attraction of the steel spring. Another little clap will be heard at station (B) in the wire spiral. By multiplying the claps and producing them in different measures you will be able as well as I am to get understood by your correspondent.*

*I am to end, Sir, and I hope, that what I said will be sufficient to have a first try; afterward you will get on quite alone.*

*I am, Sir,*

*Your most obediant Servant,*

*PH. REIS.*

*Friedrichsdorf, 13/7, 63.*

Reis's letter included the sketch here. The sketch is very similar to the (later) book illustration already shown.

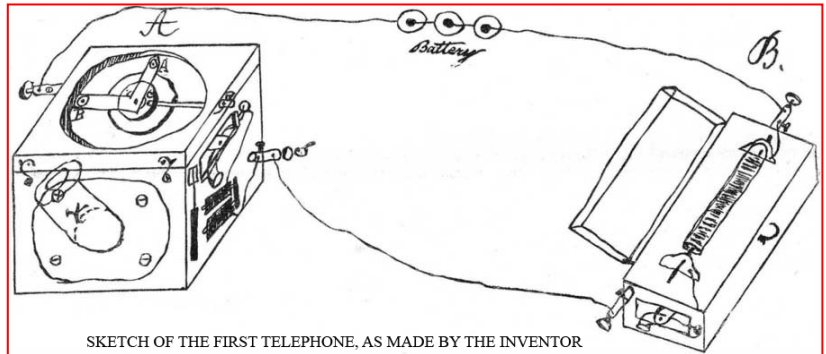
SKETCH OF THE FIRST TELEPHONE, AS MADE BY THE INVENTOR

Bell's patent success hinged on his description of his device as suitable for the transmission of speech. Could Reis's telephone transmit speech? That was the crux of the matter.

Some people looked at the mechanism and said that it couldn't. The problem seemed to be that no information on the amplitude of the signal was conveyed by the 'singing station' so the electrical signal didn't follow the fine detail of a sound waveform and hence could only poorly represent speech. Reis's device was called a telephone, almost looks like a telephone, uses electricity but doesn't quite talk like a telephone, due to an inherent flaw (as far as speech is concerned) in its method of working. The membrane transducer was described as a 'circuit breaking' instrument. It seemed to be on this basis that Bell managed to defend his patents. However, in practice Reis's telephone didn't quite work as ideally as described and it did imperfectly transmit speech.

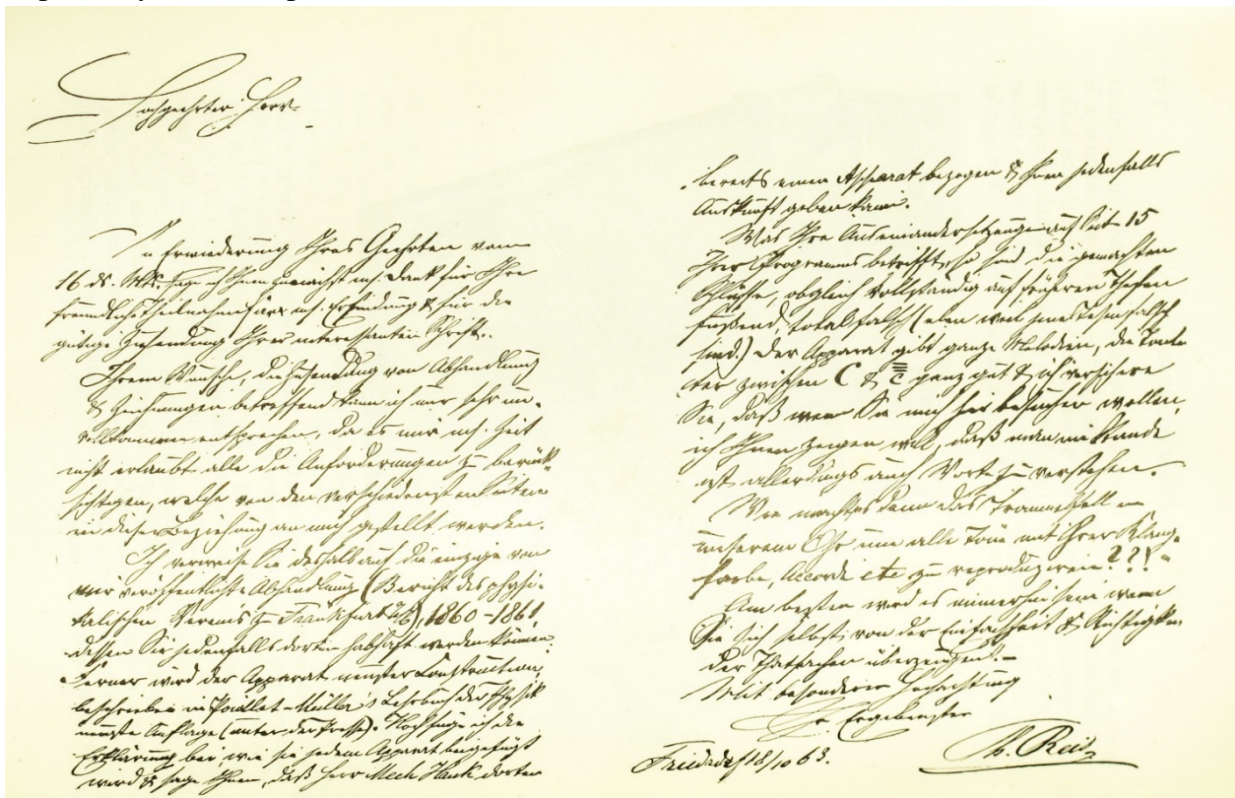

In 1863 Franz Joseph Pisko, the author of a substantial 'Lehrbuch der Physik' for schools, wrote to Reis asking for details of his 'telephon' and whether it could transmit speech. Reis's reply was included by A. R. von Urbanisky in his book that appeared in English edited by R. Wormell under the title *Electricity in the Service of Man*[7]. The autograph letter in Kurrent script is shown above.

It reads, in translation:

*Dear Sir*

*In reply to your letter of the 16th of this month I would like to first of all express my sincere thanks for your kind interest in our invention and for the gracious sending of your interesting script.*

*I can only very partially answer your request regarding the sending of treatise and drawings, as my time does not allow me to follow all the demands that are brought forward to me in this regard from various sides.*

*I therefore refer you to the only treatise that I have published (Bericht des physikalischen Vereins zu Frankfurt a/M), 1860-1861, which you will be able to acquire there. The apparatus of latest construction is further described in Pontillat-Müller's "Lehrbuch der Physik" ["Textbook of physics"], latest edition (in press). I also attach an explanation as it is enclosed with every apparatus and tell you that Mr. Mech Henk has already obtained an apparatus and will be able to provide further information.*

*Regarding your discussion on page 15 of your composition, the conclusions drawn are, while entirely consistent with previous assumptions, totally wrong (precisely because those assumptions are wrong). The apparatus reproduces entire melodies and the scale between low C and the high c''' rather well, and I assure you that if you were to visit me here, I would like to demonstrate to you that it is also possible to understand entire words.*

*What then does the eardrum in our ear do in order to reproduce all chords with their timbre, etc.??! In my opinion the best thing would be to convince yourself of the simplicity and importance of the facts.*

*Yours respectfully,*

*Ph. Reis*
*Friedrichsdorf 18.10.63*

### *Das Pferd frisst keinen Gurkensalat*

It's clear that even by 1863 Reis and others were interested in the telephone's capability of transmitting words. Sylvanus P. Thompson asserts with evidence that Reis intended the instrument to transmit speech from the beginning. In Reis's Frankfurt talk of 1861 Reis said *Hitherto it has not been possible to reproduce human speech with sufficient distinctness. The consonants are for the most part reproduced pretty distinctly, but not the vowels yet in an equal degree.* One H. F. Peter was at this Frankfurt event and he has described[8] how the demonstration went on after the meeting had broken up. Reis repeated text that was spoken to him from the remote sender. Peter said *"Philipp, you know that whole book by heart"* and he recorded *I was unwilling to believe that his experiment could be so successful unless he would repeat for me the sentences which I would give him. So I then went up into the room where stood the telephone, and purposely uttered some nonsensical sentences, for instance:"Die Sonne ist von Kupfer" (The sun is made of copper), which Reis understood as, "Die Sonne ist von Zucker" (The sun is made of sugar); "Das Pferd frisst keinen Gurkensalat" (The horse eats no cucumber salad); which Reis understood as ' Das Pferd frisst . . . . (The horse eats ... ). This was the last of these experiments which we tried. Those who were present were very greatly astonished, and were convinced that Reis's invention had opened out a great future.*

*Das Pferd frisst keinen Gurkensalat* hasn't had the historical impact of Graham Bell's *Mr. Watson, come here, I want you*, said to be the first words heard remotely through Bell's telephone in 1876. Others said that Reis's telephone just squeaked but clearly it didn't when properly adjusted.

What is beyond doubt is that Reis's phone didn't make a clear job of transmitting words and various inventors tried to improve it. The most well-known was Sylvanus P. Thompson himself, whose biography of Reis has been mentioned. In the many editions of Thompson's textbook *Elementary Lessons in Electricity and Magnetism* he repeated the claim that Reis invented the telephone[9]. Thompson also tried to improve the instrument in an effort to circumvent Bell's patent.

Had Reis had access to the then newly invented phonoautograph he would have been able to see the waveforms of his sound. For much of the 20$^{th}$ century electricians have been able to examine electrical waveforms with an oscilloscope. If Reis had had such a device, even a primitive one, he would likely have realised the problem with his 'singing station'. His transmitter did not reproduce the sound waveform correctly. Weaker sounds did vary the pressure of contact and hence produced an electrical signal that followed the sound but stronger sounds broke the circuit readily, interrupting the current. What Reis needed was the membrane to contact something that varied the electrical resistance in the circuit according to the motion of the membrane. Then the electrical signal would follow the sound waveform. Such a device was produced by Elisha Gray in 1876 but the main development of this idea had to wait for Thomas Edison a year later, following closely in Bell's footsteps.

**Origin of the Aberdeen instrument**

The Aberdeen instrument would have been bought by David Thomson, Professor of Natural Philosophy at the University of Aberdeen who had a special interest in acoustics. Indeed, he wrote the 20-page, double column article on 'Acoustics' for the first volume of the 9$^{th}$ edition of the Encyclopaedia Britannica[10] (1875), considered the acme of all editions of this work. Our instrument collection has a range of acoustic demonstration apparatus attributable to Thomson, including a Helmholtz double siren, Koenig's manometric flames, Lissajous figure tuning forks, Helmholtz resonators and even a harmonium. The article concentrated on the fundamentals of 'sound', not its application. Reis's telephone is not mentioned. By the time the 9$^{th}$ edition had reached the word 'telephone' in the 23$^{rd}$ volume, the year was 1888 and an enormous amount of telephone development had taken place. Thomas Gray wrote the seven and a half page 'telephone' article. He outlined how Reis's development drew at least implicitly on electromagnetic knowledge of the previous decades and that it had been described in over fifty publications in various countries. It's a fair conclusion to say that Reis had the concept of a telephone but the working model that swept across the globe was a development from Bell's prototypes and not from Reis's. In a way Reis was a decade too early, not from a knowledge of the necessary electrical phenomena needed to create the telephone but from a clear understanding of the nature of sound waves in speech and music.

**Reis's death**

Reis didn't live to see the telephone become one of the great inventions of the 19$^{th}$ century. From 1871 his health slowly failed with TB and in a cruel twist of fate he lost his voice. In January 1874 he died a few days after his 40$^{th}$ birthday.

In 1977 the Garnier Institute morphed into the Philipp-Reis-Schule, the largest modern comprehensive school in the district. Reis's monument in the Friedrichsdorf cemetery is shown below from the print in Sylvanus P. Thomson's biography[11].

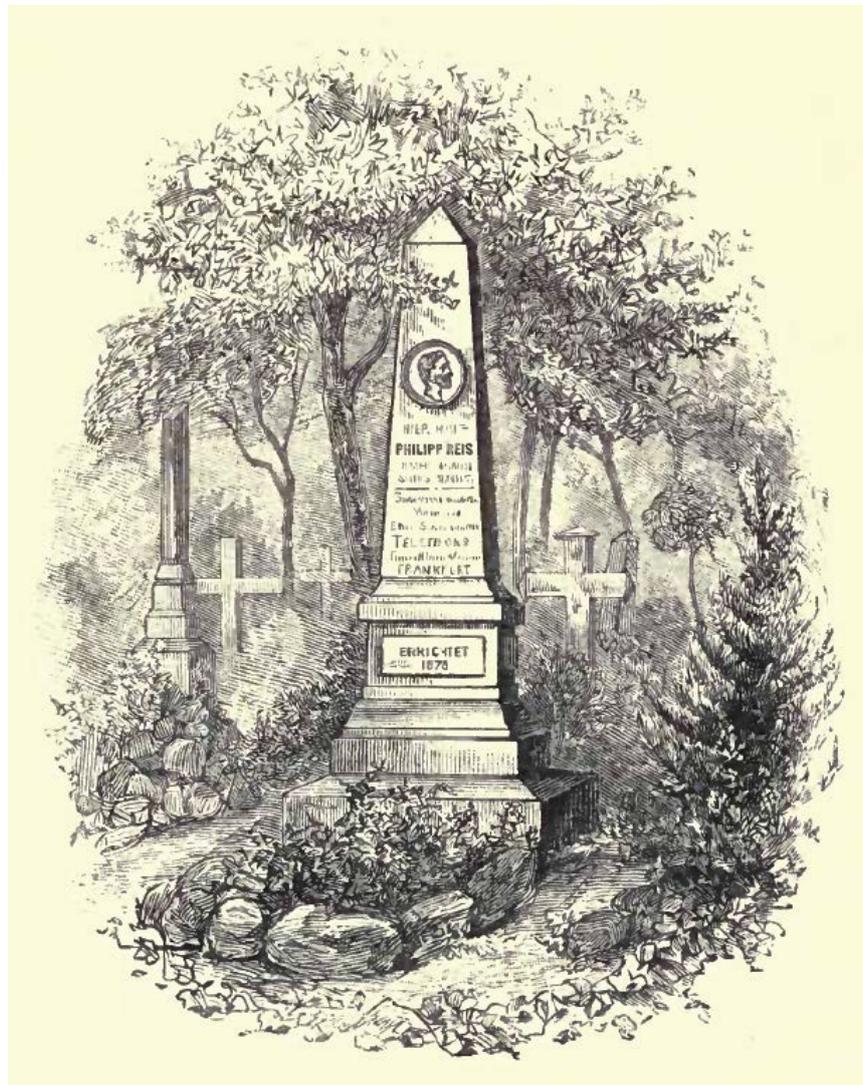

*JSR/RJHR*